\global\let\ifmypprint\iffalse 
\def\mypprint{\global\let\ifmypprint\iftrue}
\global\let\iftorefs\iffalse
\def\torefs{\global\let\iftorefs\iftrue}
\global\let\dofloatfig\iffalse
\def\floatthefig{\let\dofloatfig\iftrue}
\mypprint 
\ifmypprint
\documentstyle[prl,aps,twocolumn,epsf,rotate]{revtex}
\draft
\tighten
\floatthefig
\else
  \iftorefs 
    \catcode`\@=11 
    
    \def\figure{\let\@capwidth\columnwidth\@float{figure}}
    \let\endfigure\end@float
    \@namedef{figure*}{\let\@capwidth\textwidth\@dblfloat{figure}}
    \@namedef{endfigure*}{\end@dblfloat}
    \catcode`\@=12 
     \floatthefig 
  \fi
\fi
\input epsf.tex
\begin{document}

\newcommand{\af}{^{\{a,f\}}}
\newcommand{\Afelg}{{``A"}}
\newcommand{\A}{{\mathcal{A}}}
\newcommand{\Aus}{{A}}
\newcommand{\ba}{\begin{eqnarray}}
\newcommand{\bE}{\hat{\bf e}}
\newcommand{\beq}{\begin{equation}}
\newcommand{\bff}{{\bf f}}
\newcommand{\bit}{}
\newcommand{\bk}{{\bf k}}
\newcommand{\br}{{\bf r}}
\newcommand{\brp}{{\bf r}^{\perp}}
\newcommand{\bv}{{\bf v}}
\newcommand{\bz}{{\bf 0}}
\newcommand{\bZ}{{\bf Z}}
\newcommand{\cf}{{\it cf.}}
\newcommand{\da}[1]{\partial_{\alpha}^{#1}}
\newcommand{\dg}{\partial_{\tau}g}
\newcommand{\dt}{\partial_{\tau}}
\newcommand{\ea}{\end{eqnarray}}
\newcommand{\eeq}{\end{equation}}
\newcommand{\eg}{{\it e.g.}}
\newcommand{\eiot}{{\rm e}^{i\omega t}}
\newcommand{\en}{\eta}
\newcommand{\enut}{{``{\nut}"}}
\newcommand{\ex}{{\bf{\hat e}_x}}
\newcommand{\ey}{{\bf{\hat e}_y}}
\newcommand{\goto}{\rightarrow}
\newcommand{\gr}{{1\over\sqrt{g}}}
\newcommand{\ie}{{\rm i.e.,}}
\newcommand{\ka}{k \alpha}
\newcommand{\nut}{{\tilde\nu}}
\newcommand{\lo}{\ell_{\nut}(\omega)}
\newcommand{\n}{{\bf{\hat{n}}}}
\newcommand{\oh}{{1\over 2}}
\newcommand{\p}{\partial}
\newcommand{\ps}{\partial_{\sigma}}
\newcommand{\rg}{\sqrt{g}}
\newcommand{\rref}[1]{{Eq. \ref{#1}}}
\newcommand{\ta}{{\bf{\hat{t}}}}
\newcommand{\W}{{\cal W}}
\newcommand{\CL}{{\cal L}}
\newcommand{\Wq}{{\cal W}_q}

\twocolumn[\hsize\textwidth\columnwidth\hsize\csname @twocolumnfalse\endcsname
\title{Flexive and Propulsive Dynamics of Elastica
at Low Reynolds Numbers}
\author{Chris H. Wiggins$^{1}$ and Raymond E. Goldstein$^{2}$}
\address{$^{1}$Department of Physics, Princeton University, Princeton, NJ 08544}
\address{$^{2}$Department of Physics and Program in Applied Mathematics, 
University of Arizona, Tucson, AZ 85721}
\date{\today}
\maketitle
\begin{abstract}
A stiff one-armed swimmer in glycerine goes nowhere, but if 
its arm is elastic, exerting a restorative torque
proportional to local curvature, the swimmer can go on its way.
Considering this happy consequence and the principles of
elasticity, we study a hyperdiffusion equation for
the shape of the elastica in viscous flow, find solutions 
for impulsive or oscillatory forcing, and elucidate relevant 
aspects of propulsion.  These results have application
in a variety of physical and biological contexts, from dynamic
biopolymer bending experiments to instabilities of
elastic filaments.
\end{abstract}
\pacs{PACS numbers: 03.40.Dz, 
47.15.Gf, 
87.45.-k, 
}
\vskip1pc
]

In Stokes flow, the Aristotelian
fluid regime inhabited by the very small or the very slow,
inertia is irrelevant. 
This fact underlies the inability of a variety
of swimming motions, perfectly successful on human scales,
to generate net motion on microscopic scales \cite{Childress}. 
An oft-quoted example is
the lack of propulsion for a swimmer with only
one degree of mechanical freedom, \eg\ the paradigmatic 
scallop of Purcell's 1977 lecture which introduced many 
to the principles of Stokes flow \cite{purcell}. 
Colloquially known as ``the scallop theorem,"
this observation derives from
the more general statement that motions
invariant under $t\goto -t$ can produce no net
effect \cite{Childress}; movies of Stokes
flow must appear equally sensible when reversed \cite{movies}.

Purcell observed two ways to elude the scallop theorem:
rotate a chiral arm, or wave an elastic arm.  While the former 
dynamic is well-studied
(most notably, in the context of {\it E. coli} \cite{Lighthill}),
the latter is largely uninvestigated \cite{Machin},
despite its relation to experiments from
motility assays to dynamic studies of
biopolymer bending moduli.  To elucidate this dynamic, we here
study the motion of a one-armed swimmer with an elastic prosthesis, or 
equivalently a driven elastic filament. 
Building on  experiments showing the {\it hyperdiffusivity}
of small-amplitude planar deformations \cite{dxr}, 
we quantify how an elastic filament eludes the scallop theorem,
suggest experiments to test these results, 
and show how this analysis allows for the
measurment of bending moduli. 
Remarkably, the required mathematics \cite{us}
is central to an {\it intrinsic}
description of overdamped elastica in three dimensions.

Force-velocity proportionalities 
in Stokes flow are generally not simple;
notable exceptions are for highly symmetric objects such as the sphere
($F=6\pi\mu a v$, with $a$ the radius), and those 
for which length $L$ greatly exceeds width $d$, where 
{\it slender-body hydrodynamics} \cite{cox1,KnR} applies.
To lowest order in $1/\ln(L/d)$, 
force and velocity obey a local, anisotropic proportionality.
For velocity $v$ normal to the long axis,
the force per unit length $f=\zeta v$, with $\zeta=4\pi\mu /(\ln(L/d)+c)$,
where $c$ is an ${\cal O} (1)$ constant determined by the
shape of the object. For an elastic filament
with bending modulus $A$, $-f$ is the functional derivative of the
bending energy $A/2\int_0^L dx~ y_{xx}^2$, written here
for small planar deformations $y(x)$; thus, $f=-Ay_{xxxx}$.
At free ends the functional derivative implies boundary conditions 
of torquelessness and forcelessness:
$y_{xx}=y_{xxx}=0$.
For small deformations $v=y_t$, and with the hyperdiffusion 
constant $\nut=A/\zeta$, we have 
\beq
y_t=-\nut y_{xxxx}~,
\label{hypdif}
\eeq
perhaps the simplest model for the balance
between viscous drag and bending elasticity.

\begin{figure}
\epsfxsize=2.8truein
\centerline{\epsffile{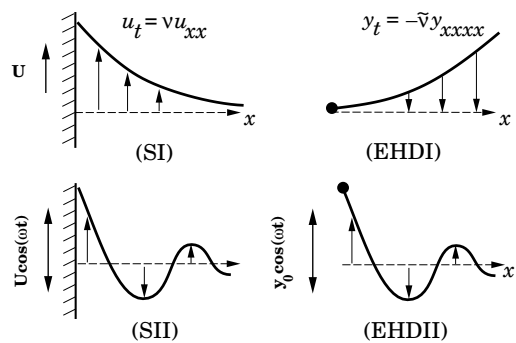}}
\bigskip
\caption{Geometry of Stokes problems I and II (left) and
of elastohydrodynamic problems I and II (right).}
\label{allfig}
\end{figure}

In 1851, Stokes suggested two problems in fluid mechanics,
here termed SI and SII (Fig. \ref{allfig}), to illustrate 
viscous diffusion of velocity \cite{Stokes}: 
SI - impulsively move a wall bounding a 
fluid; SII - oscillate the wall at frequency $\omega$.  These motivate  
two problems  
for the {\it elastohydrodynamic} equation of motion (\ref{hypdif}):
EHDI - impulsively move one end of a filament; EHDII - oscillate the end.
In SI and SII, the Navier-Stokes equation reduces to a diffusion equation
$u_t=\nu u_{xx}$, with kinematic viscosity $\nu=\mu/\rho$.
For a semi-infinite domain, the post-transient solution of SII
consists of decaying, right-moving waves,
$u\left(x,t\right)=U\exp(-\eta/\sqrt{2})
\cos\left(\eta/\sqrt{2}-\omega t\right)$,
with $\eta=x/\ell_{\nu}$, 
and viscous penetration length 
$\ell_{\nu}=(\nu/\omega)^{1/2}$.
In EHDII the analogous elastohydrodynamic penetration length is
\beq
\ell_{\nut}(\omega)=\left({\nut/\omega}\right)^{1/4}.
\label{ellnut}
\eeq
\dofloatfig
\fi

Imposing the left filament end position  $y_0\cos\omega t$
and torquelessness for the left end \cite{bead} $y_{xx}(0,t)=0$, we find
\beq
{y\over y_0}={1\over 2}\Bigl[{\rm e}^{-\tilde C \eta} 
\cos\left(\tilde S \eta + \omega t\right)
+{\rm e}^{-\tilde S \eta} \cos\left(\tilde C \eta - \omega t\right)\Bigr],
\label{ehdII}
\eeq 
where $\tilde C=\cos\left(\pi/8\right)$,  $\tilde S=\sin\left(\pi/8\right)$,
and now $\eta=x/\ell_{\nut}$.
Unlike SII, EHDII supports 
left- and right-moving waves (with different velocities and
decay lengths), despite the lack of a reflecting
right-end boundary.  

\begin{figure}
\epsfxsize=2.6truein
\centerline{\epsffile{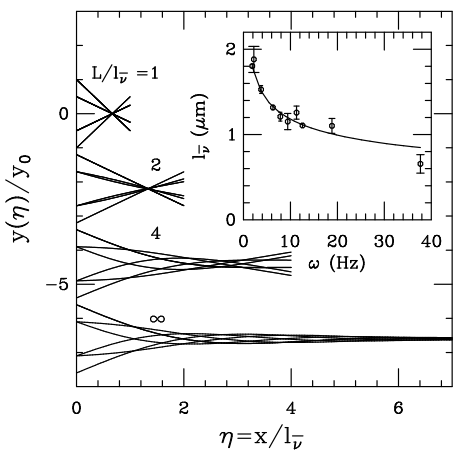}} 
\bigskip
\caption[]{Solutions to (\ref{hypdif}) for oscillatory
driving at various
rescaled lengths $\CL=L/\ell_{\nut}$.  Inset: experimental verification
\protect{\cite{dxr}} of the frequency dependence of $\ell_{\nut}(\omega)$ 
yielding a modulus $A/k_BT=7.4\mu$m, 
well within the range ($0.5 \mu$m \cite{kas} to $17\mu$m \cite{ott})
of measurements employing
statistical techniques.
} 
\label{hfig} 
\end{figure}

For finite filaments we define a rescaled length
$\CL\equiv L/\ell_{\nut}(\omega)$ and coordinate $\alpha=x/L$. 
When $\CL\lesssim 1$
the filament behaves as a rigid rod, 
while it undulates appreciably for $\CL\gg 1$. 
In this way $\ell_{\nut}$ resembles the persistence 
length $L_p$.
The exact solution
of EHDII for finite $L$ \cite{us} has an expansion in powers of
$\CL^4$ whose first terms are
\begin{eqnarray}
\label{yshort}
{y\over y_0} &\simeq&
\left(1-{3 \over 2}\alpha\right)\cos(\omega t) 
+{\CL^4 \over 1680}
\bigl(16\alpha-70\alpha^3\nonumber \\
&&+70\alpha^4-21\alpha^5\bigr)
\sin(\omega t)+{\cal O}({\cal L}^8)~.
\end{eqnarray}
At order $\CL^0$, the filament is a straight 
rod that interestingly pivots about a point two-thirds of
its length.  Flexive corrections at ${\cal O}(\CL^4)$ break
time-reversal invariance.
Solutions of increasing $\CL$
are shown in Fig. \ref{hfig}, whose inset
shows the results of an experiment on actin \cite{dxr} in which 
observed shapes
in a range of frequencies were fit for $\ell_{\nut}$
to the exact expressions \cite{us}, verifying the
scaling $\ell_{\nut}\sim \omega^{-1/4}$
as well as providing a novel dynamic technique for measuring
the bending modulus $A$.

\begin{figure}[tb]
\epsfxsize=2.8truein
\centerline{\epsffile{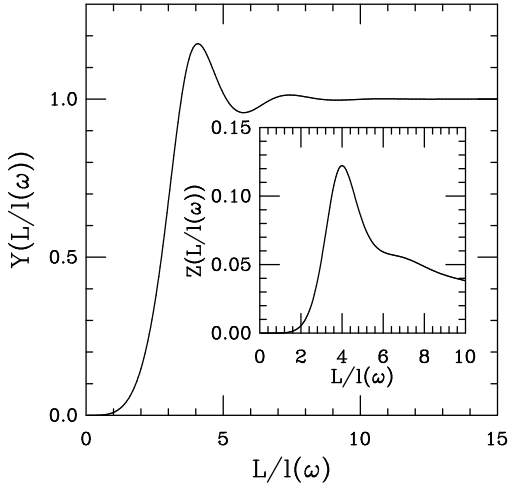}}
\smallskip
\caption[test2]{Scaling functions $Y$, for propulsive force,
and $Z$, for efficiency, versus rescaled length $\CL=L/\lo$.}
\label{upsfig}
\end{figure}

The propulsive force $F_x$ imparted 
to the fluid by the filament (or vice versa) 
may be computed by integrating the
projected  elastic force density ${\bf f}\cdot \hat{\bf e}_x$
along the filament.
Interestingly, with the boundary conditions of EHDII
and the geometrically
exact form of ${\bf f}$ this result is expressible 
\dofloatfig
\fi
\noindent in terms of the curvature $\kappa$ and tangent angle $\theta$
{\it at the forcing point}: 
$F_x=A\kappa_s\sin\theta(s=0)\simeq Ay_xy_{xxx}(x=0)$.
The time average of this quantity over one period gives
\beq
\bar F_x=
{y_0^2\zeta\vert\omega\vert\over 4\sqrt{2}}
Y\left(\CL\right)\label{feqn}~
\eeq
For $\CL\ll 1$,  
$Y\simeq(11/3360)\CL^4$, 
so $\bar F_x\sim y_0^2\zeta^2\omega^2L^4/A$; 
a short (or infinitely stiff) pivoting filament 
produces no net force (by the scallop theorem).  Flexibility leads to
a net leftward propulsion, as the right-moving
waves 
dominate the left-movers.
An unexpected, fascinating feature (Fig. \ref{upsfig})
is the maximum in $Y$ indicating an optimal value of the 
length $L^{*}\simeq 4.07(A/\zeta\omega)^{1/4}$.

In a familiar way, this force 
is associated with the trajectory of the filament shape
in a low-dimensional projection of 
configuration space.  The relation
$F(t)=Ay_x(0,t)y_{xxx}(0,t)$ and
the equation of motion imply
\beq
\bar F_x={\zeta\omega\over 2\pi}
\int_0^{2\pi/\omega}\!\! dt y_x(0,t){\partial\over \partial t}\int_0^L\!\! dx y
={\zeta\omega\over 2\pi}\oint \theta_0 d\A~,
\eeq
where $\A$ is the area under the curve $y(x)$ and $\theta_0\simeq y_x$ is
the tangent angle at the left end. 
Thus, the net force during the cyclic motion is the area enclosed in the associated `Carnot diagram' in
$\A - \theta_0$ space;
it results from pushing aside some volume
of fluid (area, in two dimensions), projected via $\theta_0$
in the direction of propulsion.
For EHDII,
the trajectory is an ellipse; it thins to a straight
line for the time-reversible pivoting of a rod, encloses no area, 
and thus produces no force.
Observe the intuitive result that net propulsion in the
transverse direction, proportional to $\oint d\A=0$, vanishes
identically.

We estimate the efficiency ${\cal E}$ of
this motion \cite{Childress} by comparing
the power $P_\parallel=\bar F_x v_x\simeq \bar F_x^2/L\zeta_{\parallel}$ 
for longitudinal propulsion
to the power $P_\perp=F_y v_y\simeq\int ds\zeta_{\perp}y_t^2$ dissipated
in transverse motions, to obtain
\beq
{\cal E}=\left({y_0\over 2 \ell_{\nut}}\right)^2{\zeta_\perp\over\zeta_\parallel} 
Z\left(\CL\right)~,
\label{efficiency}
\eeq
where $Z(\CL)$ is 
the scaling function shown in the inset
to Fig. \ref{upsfig}.  
Filaments short relative to $\ell_\nut(\omega)$ flex little
and produce little propulsion, while long ones have 
excess drag from the nearly straight regions far away from the point of
forcing, yielding a sharp maximum at $\CL\sim 4.0$.

These observations suggest experiments in the spirit of
those by Taylor \cite{Taylor} on swimming
with a helical flagellum.  
Exploiting the results of EHDII,
perhaps carried out on microfilaments with laser \cite{dxr} or
magnetic tweezers, or on macroscopic objects, 
one might measure the propulsive force through the transverse displacement 
at the forcing point, test for the predicted maximum as a function of 
frequency, investigate the role of nonlinearities, and study interactions 
between flexing filaments. Analogous experiments incorporating 
twist (perhaps via magnetic
optically-trapped beads, as in \cite{strick}) could elucidate
instabilities exhibited by
helical motion of flexible filaments \cite{phil} and
associated propulsion.
\dofloatfig
\fi

A swimmer's arm may also move by discrete, noncyclic strokes. 
The associated model dynamic (EHDI) is the relaxation of
a filament to an equilibrium shape given some 
boundary condition at the point of attachment.
Since (\ref{hypdif}) is linear, 
we may subtract from its general solution a particular
solution consistent with nonzero boundary conditions or 
external driving to obtain a 
homogeneous problem with $0$-valued boundary conditions.
This motivates the construction of a self-adjoint operator
from ${\cal H}\equiv \partial_x^4$, whose 
well-known eigenfunctions are 
\begin{eqnarray}
\Wq(x)&=&
a_1\sin\left(qx\right)+
a_2\cos\left(qx\right)\nonumber \\
&&+a_3\sinh\left(qx\right)+
a_4\cosh\left(qx\right)~,
\label{biharmonic}
\end{eqnarray}
where ${\cal H}\Wq=q^4\Wq$.
The ten distinct 
choices of boundary conditions for which $\cal H$ is self adjoint \cite{us}
determine the distinct coefficients $\{a_i\}$ of $\Wq$ and 
the allowed wavenumbers $q$.  These $q$ are roots of transendental
solvability conditions;
certain boundary conditions
are satisfied by Fourier modes (those for which $a_3=a_4=0$).

The completeness of the set $\Wq$
yields the homogeneous solution to (\ref{hypdif}),
\beq
y(x,t)=\sum_q
\Wq(x){\rm e}^{-\nut q^4 t}\int_0^L dx\Wq (x)y(x,0)~,
\label{project}
\eeq
relevant to a second class of techniques for 
measuring bending moduli. One 
recent example \cite{Felg} is
the relaxation of the free end of an initially bent microtubule whose opposite
end 
is clamped to a cover slip via the axoneme from which it is nucleated.
Using typical material parameters
for microtubules, ($L\sim \mu$m, $\zeta\sim$ cP, $L_p\sim$ mm), 
we observe from (\ref{project}) that beyond the inital few milliseconds
the filament straightens with a time constant $\tau_1=1/\nut q_1^4$;
here $q_1\simeq 1.875/L$ the smallest root of the clamped-free solvability
condition $\cos(qL)\cosh(qL)=-1$ \cite{Gittes}. Measurement of $\tau_1$ gives $A$.

Finally, we apply these results to an important
subject in pattern formation:
{\it intrinsic} formulations
of filament motion \cite{Ueda} and their extention to nonplanar
 dynamics.
The dynamical evolution of the Frenet-Serret (FS) curvature $\kappa$ and torsion $\tau$ are 
singular at inflection points (where $\kappa=0$ and $\tau$ is undefined), a
problem removed by instead using the parameterization \cite{GnL,hasimoto}
\beq
\psi(s,t)=\kappa(s,t) {\rm e}^{i\phi(s,t)}~, \ \ \ \ 
{\bf w}=(\hat{\bf n}+i\hat{\bf b}){\rm e}^{i\phi}~,
\label{psi}
\eeq
where $\phi(s,t)=\int^s\!\! ds' \tau(s',t)$, and $\hat{\bf n}$ and $\hat {\bf b}$
are the normal and binormal vectors. 
Across an inflection point at, say, $s=0$, 
the normal and binormal vectors $\hat{\bf n}$
and $\hat{\bf b}$ flip sign, the
torsion has a singular piece, $\tau=\pi\delta(s)$, so
there is a discontinuity $\phi(0^+)
-\phi(0^-)=\pi$.  Yet, this leaves ${\bf w}$ and $\psi$ continuous. 
The curve ${\bf r}(s)$ is constructed from $\psi$ 
via new FS equations: 
${\bf w}_s=-\psi\hat{\bf t}; 
\ \hat{\bf t}_s={\rm Re}(\psi^*{\bf w})$.
Some elementary shapes have very simple
$\psi$-representations: the straight line $\psi=0$,
the circle $\psi=a$, the helix
$\psi=a{\rm e}^{ips}$ ($a,p$ constant).

\begin{figure}
\epsfxsize=2.8truein
\centerline{\epsffile{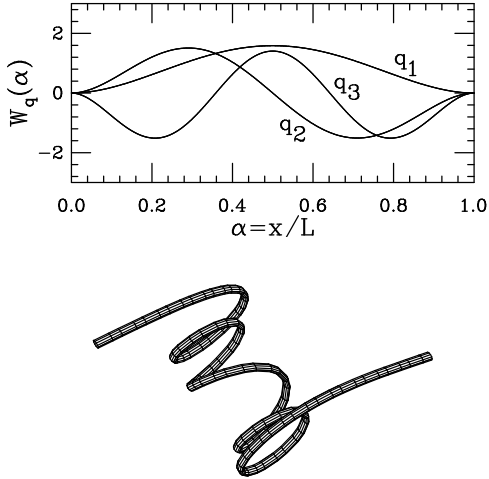}}
\smallskip
\caption{Biharmonic eigenfunctions and a nonplanar elastica.
Top: ${\cal W}_q$ with boundary conditions ${\cal W}=\partial_{\alpha}{\cal W}=0$.
Bottom: free filament with $\psi=15({\cal W}_{q_1}+i{\cal W}_{q_2})$.}
\label{hasimotofig}
\end{figure}

The intrinsic inextensible evolution for $\psi$ \cite{GnL},
\beq
\psi_t=(\partial_{ss}+\vert\psi\vert^2)\Upsilon + \psi {\rm Im}
\int^s\!\! ds' \psi_s\Upsilon^*+W\psi_s~,
\label{psi_int}
\eeq
depends upon the velocity components ${\bf r}_t=U\hat{\bf n}
+ V\hat {\bf b}+W\hat{\bf t}$ through the complex velocity
$\Upsilon=(U+iV){\rm e}^{i\phi}$.  It can be shown \cite{filaments} from the
governing equations for viscous motion of
filaments, 
that $\Upsilon= -\nut(\psi_{ss}+(1/2)\vert\psi\vert^2\psi)$.  Thus,
the $\psi$ dynamics is
hyperdiffusive as in (\ref{hypdif}) \cite{GnL},
\beq
\psi_t \simeq -\nut\psi_{ssss}+\cdots ~.
\label{beam_psi}
\eeq
For an elastic energy
${\cal E}=(A/2)\int\! ds \kappa^2$ the boundary conditions 
at free ends are $\kappa=\kappa_s=\kappa\tau=0$, or simply 
$\psi=\psi_s=0$. 
Since the self-adjoint operator
is again ${\cal H}=\partial_s^4$, the ${\cal W}_q(s)$,
now functions of {\it arclength}, are the relevant eigenfunctions.  
Thus, the $\psi$ formulation, the natural singularity-free {\it mathematical}
representation of the curve, also compactly
encodes the {\it physical} boundary conditions and dynamics.  Since the
${\cal W}_q$ form a complete set, the evolving shape of
{\it any} filament is expressible as
\[
\psi(s,t)=\sum_q c_q(t){\cal W}_q(s)~,
\]
the time evolution being a nonlinear dynamical system in
$c_q(t)\equiv\langle\Wq\vert\psi(s,t)\rangle$; in the linearized regime, 
$c_q(t)=c_q(0)\exp(-\nut q^4 t)$.  The modes $\Wq$ are like 
the Rouse modes of a flexible polymer,
and the ``clamped" eigenfunctions of $y_t=-\nut y_{xxxx}$ 
(with $y=y_x=0$ at $x=0,L$)
are the ``free" eigenfunctions of $\psi_t=-\nut\psi_{ssss}$, 
with $\cos(qL)\cosh(qL)=1$.  The first three are shown in
Fig. \ref{hasimotofig}. When the complex coefficients $c_q$ share a common 
phase the curve is planar, otherwise it has torsion.  The latter case
is shown in Fig. \ref{hasimotofig} as a coiled three-dimensional
filament composed of the first two modes, reconstructed from
the complex FS equations.  The simplification achieved with the $\psi$
formulation is clear.
\dofloatfig
\fi

This formalism easily generalizes to {\it twisted} filaments
\cite{filaments}, as found for instance in bacterial systems that
undergo hierarchical buckling and writhing instabilities \cite{neil}.
For a constant twist density $\Omega$ and twist elastic constant $C$
the moment relation ${\bf M}=A\kappa\hat{\bf b}
+C\Omega\hat{\bf t}$ can be shown to yield the complex velocity \cite{Shi}
\beq
\Upsilon=-\nut(\psi_{ss}+{1\over 2}\vert\psi\vert^2\psi)
+ i\nut C\Omega\psi_s~,
\label{ups_twist}
\eeq
so the intrinsic dynamics (\ref{beam_psi})
is explicitly complex 
\beq
\psi_t \simeq -\nut(\psi_{ssss}+i\Omega\Gamma\psi_{sss}) +\cdots ~,
\label{beam_psi_twist}
\eeq
with $\Gamma=C/A$.  A helical perturbation 
$\psi\sim \exp(iks+\sigma t)$ to a straight filament 
leads to a growth rate $\sigma= -\nut(k^4
+\Gamma\Omega k^3)$.  This describes a writhing instability on a length
scale $L_{Wr}= 2\pi/\Omega\Gamma$ to a helix whose handedness
is set by the sign of the twist density $\Omega$.  A stability
analysis \cite{filaments} about a loop of radius $R$ yields 
not only the criterion $\Gamma\Omega R=\sqrt{3}$ for the onset of the
primary instability \cite{Zajac}, but also the {\it growth rates}
$\sigma_{\pm}$ for the coupled bending
and writhing modes.
Extensions to the fully nonlinear regime are
then straightforward, leading to a ``dynamics of
twist and writhe" that complements important existing Hamiltonian
formulations \cite{Goriely}.

We thank A. Ott and D.X. Riveline for collaborations,
S. Childress, J. Kessler, and  C. O'Hern for discussions and
R. Kamien, P. Nelson, and T. Powers for 
insightful and instructive protests.
This work was supported by an NSF Presidential Faculty Fellowship,
DMR 96-96257 (REG).
This paper is dedicated to the late E. Purcell.

\end{document}
\end